\documentclass[twocolumn,showpacs,superscriptaddress,amsmath,amssymb,prl]{revtex4-1}
\usepackage[normalem]{ulem}
\usepackage{epsfig}
\usepackage{amsfonts}
\usepackage{amsmath}
\usepackage{slashed}
\usepackage{graphicx}
\usepackage{color}
\usepackage{mathtools}
\usepackage{subfigure}
\usepackage{graphicx}

\usepackage{stmaryrd}
\usepackage{amssymb,psfrag}
\usepackage[normalem]{ulem}
\usepackage{caption}

\newcommand{\beq}{\begin{eqnarray}}
\newcommand{\eeq}{\end{eqnarray}}

\newcommand{\nn}{\nonumber}

\begin{document}
\title{Exploring Nucleon Structure through Sub- and Near-threshold $\phi$-Meson Photoproduction}

\author{Wei Wang}
\email{wei.wang@sjtu.edu.cn}
\affiliation{INPAC, Key Laboratory for Particle Astrophysics and Cosmology (MOE), Shanghai Key Laboratory for Particle Physics and Cosmology,
School of Physics and Astronomy, Shanghai Jiao Tong University, Shanghai 200240, China}

\author{Ji Xu}
\email{xuji@lzu.edu.cn}
\affiliation{Frontiers Science Center for Rare Isotopes, and School of Nuclear Science and Technology, Lanzhou University, Lanzhou 730000, China}

\author{Xing-Hua Yang}
\email{yangxinghua@sdut.edu.cn}
\affiliation{School of Physics and Optoelectronic Engineering, Shandong University of Technology, Zibo, Shandong 255000, China}

\author{Ya-Teng Zhang}
\email{zhangyateng@zzu.edu.cn}
\affiliation{School of Physics and Microelectronics, Zhengzhou University, Zhengzhou, Henan 450001, China}

\author{Shuai Zhao}
\email{zhaos@tju.edu.cn}
\affiliation{School of Science, Tianjin University, Tianjin 300072, China}

\date{\today}

%\date{\today}
\begin{abstract}
We propose to investigate short-range correlations (SRCs) in nuclei by studying sub and near-threshold photoproduction of $\phi$-meson. The feasibility of exploring this nuclear structure in electron-positron collision experiments such as Beijing Spectrometer III (BESIII) and Super $\tau$-Charm Facility (STCF) was investigated by utilizing their beryllium-based ($^{9}$Be) beam pipe. The cross sections for these processes as well as the likelihood of detection by BESIII and STCF are calculated. Due to the limited number of events, BESIII is inadequate for the proposed measurement, while the future STCF provides a higher probability of successfully measuring this process. These findings suggest that electron-positron collider could play a valuable role in elucidating the fundamental physics behind the nuclear modification of parton distribution functions. This proposed analysis of photon-nucleon interactions represents uncharted territory, promising fresh prospects for applications in both particle and nuclear physics.
\end{abstract}

\maketitle

\textit{Introduction.}---The strong interaction between nucleons (protons and neutrons) serves as an essential binding force that upholds the structural integrity of the atomic nucleus. This force arises from the intricate interplay among quarks and gluons, the fundamental constituents of nucleons, and is governed by the principles of quantum chromodynamics (QCD). On experimental side, the study of internal structure of matter  often relies on scattering experiments such as  high-energy electron scattering measurements, which play a particularly vital role such as  in the identification of short-range correlations (SRCs) \cite{Frankfurt:1988nt}.

When nucleons are bound into atomic nuclei, they are close enough to experience significant attraction or repulsion from each other.  These strong interactions can result in hard collisions between nucleons, leading to the creation of pairs of highly energetic and off-shell nucleons. This phenomenon, characterized by nucleon pairs with large relative momentum and small center-of-mass (c.m.) momentum compared to the single-nucleon Fermi momentum, is known as SRCs. SRCs are a significant yet not fully understood aspect of nuclear structure, and  understanding their role in the atomic nucleus is crucial for modeling various nuclear, particle, and even astrophysics phenomena \cite{Hen:2016kwk,Arrington:2022sov}. Following the observation of SRCs \cite{Frankfurt:1993sp}, it was initially believed that their contribution scaled with nuclear density \cite{CLAS:2005ola}. However, the measurement of the scaling factor $a_2^A$ for $^9$Be in \cite{Fomin:2011ng} revealed that clustering effects, i.e., the specifics of nuclear structure, significantly impact the relative contribution of SRCs. Recent studies also suggest a strong connection between SRCs and the well-known EMC effect \cite{EuropeanMuon:1983wih,Weinstein:2010rt,Hen:2012fm}, offering a new perspective for investigating the origins of EMC. To strengthen the physical argument for this connection, a rigorous examination is urgently needed, particularly focusing on gluonic processes.

Current research primarily focuses on the quark sector of the partonic structure within nuclei, but it is crucial to also investigate SRC contributions in the gluonic sector. Short-range nucleon pairs, which exhibit a denser local environment and higher momenta with significant virtuality, are closely tied to strong interactions. To gain deeper insights, it is important to explore the distributions of gluons, the mediators of strong interactions, within nuclei. The investigation of gluonic probes to SRCs can be done through heavy flavor production in deep inelastic scattering (DIS)~ \cite{Aschenauer:2017oxs,Xu:2019wso,Hatta:2019ocp,Sun:2021pyw}. Recent studies have examined gluon parton distribution functions in nuclear environments and proposed a linear relationship between the magnitude of the gluon EMC effect and the cross section of sub-threshold photoproduction of $J/\psi$~\cite{Wang:2024cpx}, a process dominated by gluonic SRC contributions. However, our current knowledge in this area is limited. Early attempts at sub-threshold $J/\psi$ production from a carbon target at Jefferson Laboratory did not yield any observed events~\cite{Bosted:2008mn}. In Ref.\,\cite{Pybus:2024ifi}, a new measurement of $J/\psi$ photoproduction from nuclei, including one data point below the energy threshold of $8.2$\,GeV has been presented. Other experiments have mainly focused on near-threshold photoproduction of $J/\psi$~\cite{GlueX:2019mkq,GlueX:2023pev}, which can be used to search for pentaquark candidates. This highlights the necessity for further theoretical and experimental investigations to probe the distribution of gluons bounded in SRC pairs.

The exploration of the sub- and near-threshold regime of strangeness production offers a new perspective on this subject. The $\phi$ meson, being a conventional $s \bar s$ state with a narrow width of $4.3 \,\textrm{MeV}$, can act as a ``gluon filter''. Its $s \bar s$ nature suppresses light quark  contamination, and its photoproduction dynamics is governed by gluon exchange \cite{Hatta:2025vhs}. Therefore, by analyzing this production data in the near-threshold (and especially in the sub-threshold) regime, we can investigate the impact of SRCs on the gluon distributions in the nucleon. In the literature, photoproduction of the $\phi$ meson has attracted significant interest as it has the potential to shed light on a wide range of physics phenomena. This process can be utilized to explore the restoration of chiral symmetry in dense nuclear matter, test the pomeron exchange mechanism, and quantify the nuclear medium modification of vector meson properties \cite{Qian:2010rr,Wang:2022uch,Gao:2000az,Sekihara:2010rw}. Additionally, studies suggest that the nuclear transparency ratio of $\phi$-meson photoproduction is sensitive to the details of the nuclear structure. Early experimental data of $\phi$ photoproduction are generally sparse, characterized by wide energy bin-widths and limited statistical precision \cite{Aachen-Berlin-Bonn-Hamburg-Heidelberg-Munich:1968rzt,Mcclellan:1971tk,Anderson:1972ac,Egloff:1979mg,ZEUS:1999ptu}, and more recent findings originate from measurements near the threshold by LEPS and CLAS \cite{LEPS:2005hax,LEPS:2009nuw,Dey:2014tfa,CLAS:2007xhu,Chang:2007fc}.

In this work, our goal is to investigate the potential of sub- and near-threshold $\phi$-meson photoproduction to probe nucleon structure, with a particular interest in the electron-positron collision experiments, such as the Beijing Spectrometer III (BESIII) and the Super $\tau$-Charm Facility (STCF) \cite{BESIII:2009fln, Achasov:2023gey}. We aim to leverage the continuous photon beam, with its feasible energies and high luminosity at these facilities, to their full potential. To estimate the cross section for sub-threshold production at BESIII, we will utilize the energy fraction parameter model \cite{Brodsky:2000zc}. Additionally, we will explore the effective luminosity of the photon flux produced by $J/\psi (\psi(3686))$ decays, radiative Bhabha and $e^+ e^-\to \pi^+\pi^- \gamma_{\text{ISR}}$, considering the distribution of target materials within the experimental setup at BESIII. Based on our results, we find that the data of BESIII would be inadequate, whereas the future STCF provides a higher probability of successfully measuring this process  due to its high luminosity to explore the nuclear structure through $\phi$-meson photoproduction. Future experimental efforts to validate this proposal have the potential to advance our understanding of the nucleon substructure significantly.  In addition to exploring SRCs, we note that the interactions between photons and materials in the beam pipe at an electron-positron collider provide opportunities for a diverse range of high-precision physics measurements. This encompasses research areas such as the $\phi-N$ bound states, the strangeness content of the nucleon, and the proton mass radius.

%%%%%%%%%%%%%%%%%%%%%%%%%%%%%%%%%%%%%%%%%%%%%%%%%%%%

\textit{The sub- and near-threshold $\phi$-meson photoproduction cross section.}---We now turn to the theoretical framework for $\phi$ photoproduction. There exist some experimental data in the near-threshold region, which can be compared with theoretical models. For the sub-threshold case, a photon with energy $E_\gamma \leq 1.57 \,\textrm{GeV}$ hits the beam pipe. The production of $\phi$-meson is kinematically forbidden for this photon interacting with a stationary nucleon, but is accessible as the nucleon momentum increases, providing a way to isolate scattering from moving nucleons and thus study high-momentum nucleons in SRCs. Although the $\phi$ is commonly observed experimentally, the underlying interaction mechanism is not well understood and depends on the energy of the incoming photon.

Here, we consider the process $\gamma  N \to \phi N$ in the region from sub-threshold $E_\gamma \simeq 1.30 \textrm{GeV} $ to near-threshold $E_\gamma \simeq 2.30 \,\textrm{GeV}$ in the lab frame. The energy fraction parameter model is adopted \cite{Brodsky:2000zc}
\begin{eqnarray}
  \chi_\gamma = \frac{M_{\phi}^2}{2E_\gamma M_N}+\frac{M_\phi}{E_\gamma} \,.
\end{eqnarray}
The threshold limit corresponds to $\chi_\gamma \to 1$. Therefore, we can apply a simple power law of $(1-\chi_\gamma)$ for the near-threshold $\phi$ production,
\begin{eqnarray}\label{model1}
  \sigma_{\gamma N \rightarrow \phi}\left(W_{\gamma N}\right)=\sigma_0^{\gamma N}\left(1-\chi_\gamma\right)^\beta \,,
\end{eqnarray}
where $\sigma_0^{\gamma N}$ and $\beta$ are two parameters representing threshold behavior.

For the case of $\gamma + (p n) \to \phi $, where the photon interacts with a nucleon in SRC pair, we argue that the basic formula holds, and one only needs to modify $\chi_\gamma$ \cite{Xu:2019wso},
\begin{eqnarray}
  \chi_\gamma \rightarrow \tilde{\chi}_\gamma=\frac{M_{\phi}^2}{2 E_\gamma 2 M_N}+\frac{M_{\phi}}{E_\gamma} \,,
\end{eqnarray}
where we have used two-nucleon mass to represent the threshold kinematics. Therefore, the SRC cross section can be written in the same form,
\begin{eqnarray}
  \sigma_{\gamma(p n) \rightarrow \phi}\left(W_{\gamma p}\right)=\sigma_0^{\gamma(p n)}\left(1-\tilde{\chi}_\gamma\right)^{\beta_2} \,.
\end{eqnarray}
At higher energy $(1-\tilde{\chi}_\gamma) \to (1-\chi_\gamma)$, so we further assume that $\sigma_0^{\gamma(p n)}=2\sigma_0^{\gamma N}$. For the threshold behavior, $\sigma_{\gamma(p n) \rightarrow \phi}$ may be different from $\sigma_{\gamma N \rightarrow \phi}$, we estimate the contributions across a broad range of $\beta_2 = n_2 \beta$, with $n_2$ ranging from 1 to 3.

Therefore, we have the following expression for $\phi$ production in $\gamma A$ process,
\begin{eqnarray}\label{energyregion1}
  && \sigma_{\gamma A \to \phi}(W_{\gamma N})/A = \sigma_{\gamma N \rightarrow \phi}\left(W_{\gamma N}\right) \nn\\
  && \quad + \frac{n_{\textrm{SRC}}^A}{A} \Big[ \sigma_{\gamma(p n) \rightarrow \phi}\left(W_{\gamma N}\right) -2\sigma_{\gamma N \rightarrow \phi}\left(W_{\gamma N}\right)  \Big] \,.
\end{eqnarray}
Here $n_{\textrm{SRC}}^A$ is the number of $np$-SRC pairs in nucleus $A$. We do not take into account the three-body SRCs and the isospin symmetry is applied to simplify the cross section calculations.

When the energy of photon is below the threshold, $E_\gamma \!=\! 1.57$\,GeV, i.e. $W_{\gamma N}<M_{N}+M_\phi$, the only contribution comes from the $\gamma (pn)$-term in Eq.\,(\ref{energyregion1}),
\begin{eqnarray}\label{energyregion2}
  \sigma_{\gamma A \to \phi}(W_{\gamma N}\!<\!M_{N}\!+\!M_\phi)/A = \frac{n_{\textrm{SRC}}^A}{A} \sigma_{\gamma(p n) \rightarrow \phi}\left(W_{\gamma p}\right) .
\end{eqnarray}

\begin{figure}[htbp]
%\begin{center}
\includegraphics[width=1\columnwidth]{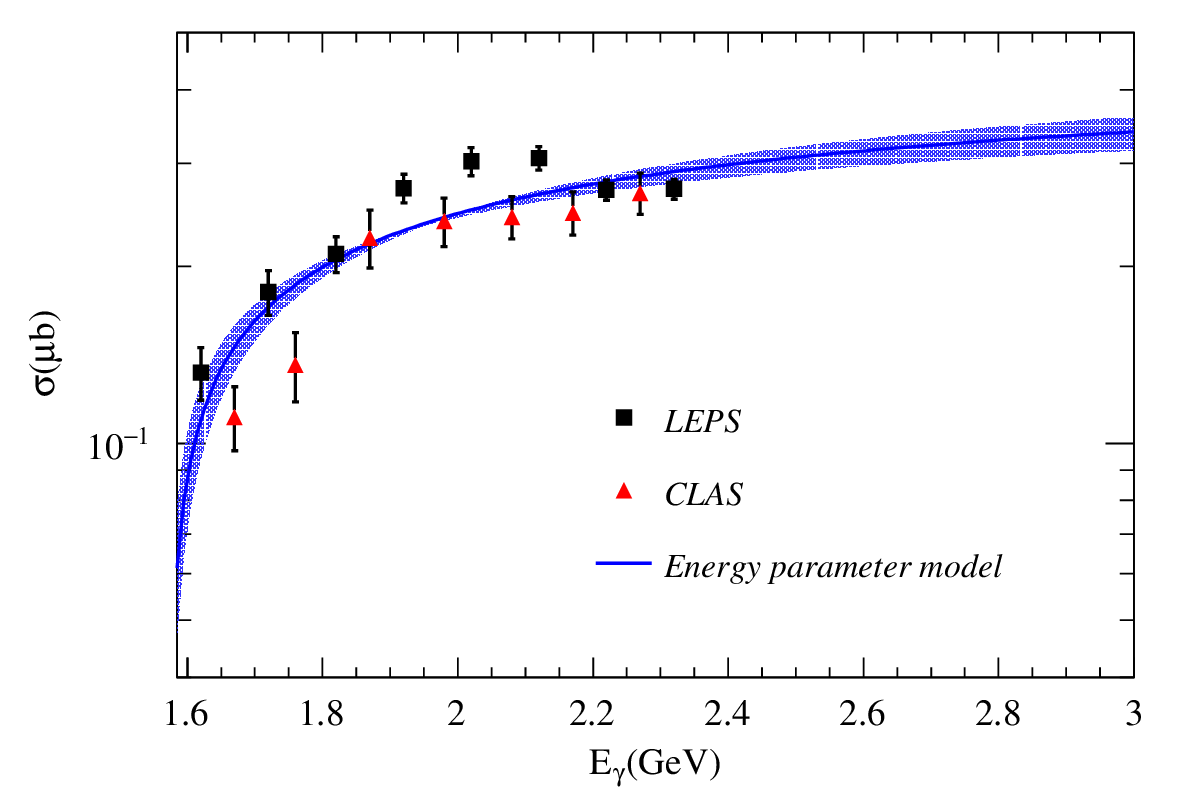}
%\end{center}
\caption{The near-threshold cross section of $\phi$-meson photoproduction as a function of incoming photon energy $E_\gamma$. Data from LEPS \cite{LEPS:2005hax} and CLAS \cite{Dey:2014tfa} are used in the fit. The solid-line (blue) is the energy fraction parameter model, with the fitted parameters $\sigma_0^{\gamma N} = 0.456 \pm 0.054 \,\rm{\mu b}$ and $\beta = 0.398 \pm 0.072$.  }
\label{LEPS1}
\end{figure}

Afterwards, we will ascertain the value of $\sigma_0^{\gamma N}$ and $\beta$ in Eq.\,(\ref{model1}) from data in LEPS \cite{LEPS:2005hax} and CLAS \cite{Dey:2014tfa} experiments near-threshold, where the energy of photons lies in the range $1.57 \!\sim\! 2.37$ GeV and $1.63 \!\sim\! 2.34$ GeV, respectively. Fig.\,\ref{LEPS1} shows the cross section of $\phi$-meson photoproduction predicted by the energy fraction parameter model at the near-threshold, which agrees well with the experimental data. The determined parameters are $\sigma_0^{\gamma N} = 0.456 \pm 0.054 \,\rm{\mu b}$ and $\beta = 0.398 \pm 0.072$. Besides, this result is also consistent with the two gluon exchange and the pomeron models near the threshold region. Based on these facts, we consider the cross section of $\phi$-meson photoproduction can be reliably predicted.

\begin{figure}[htbp]
%\begin{center}
\includegraphics[width=1\columnwidth]{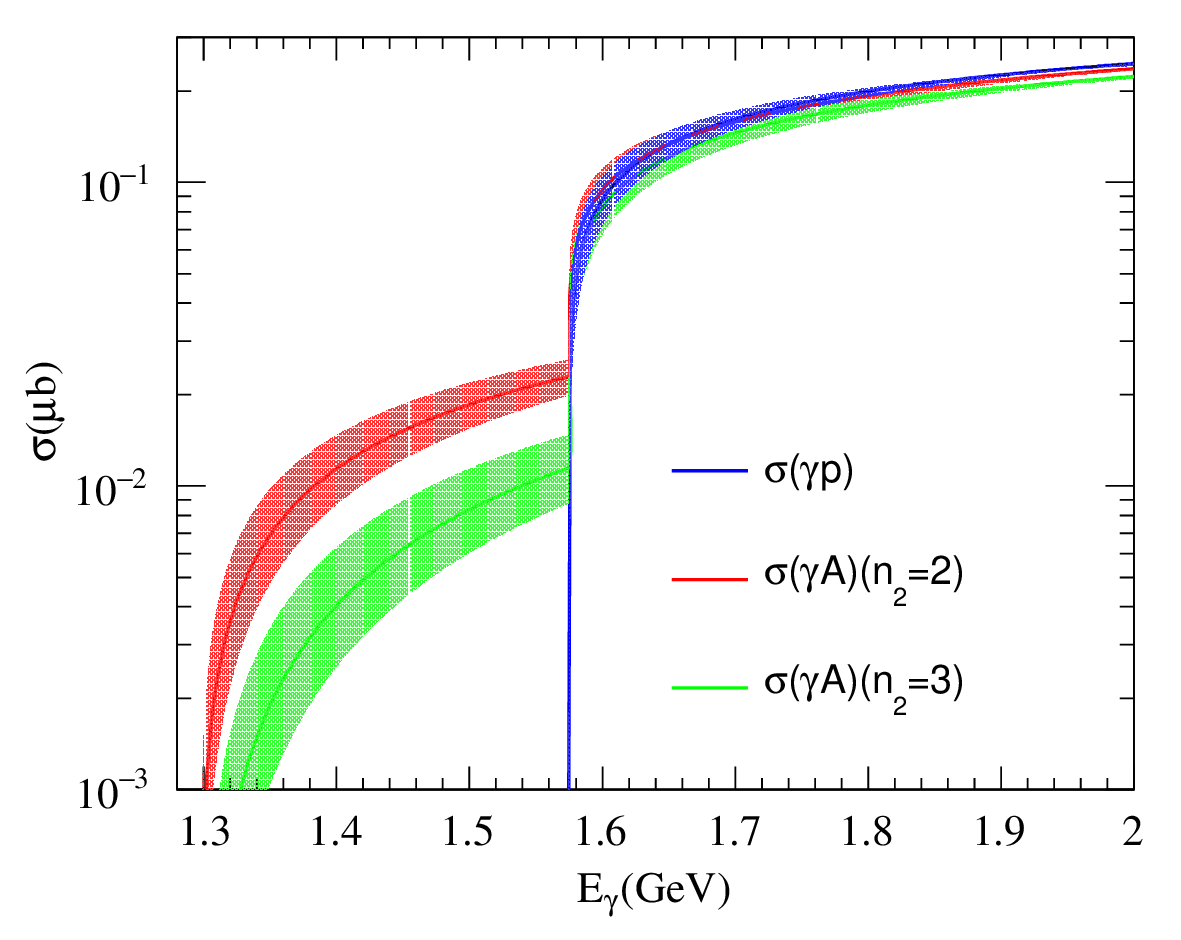}
%\end{center}
\caption{Sub- and near-threshold $\phi$-meson photoproduction cross sections in $\gamma A$ collision, the interaction with free proton is also presented as comparison. }
\label{M11}
\end{figure}

In Fig.\,\ref{M11}, we plot the per-nucleon sub- and near-threshold $\phi$ production cross section in $\gamma A$ collision, based on Eqs.\,(\ref{energyregion1}) and (\ref{energyregion2}). $(n_{\textrm{SRC}}^A / A) \sim 10\%$ is adopted according to \cite{Subedi:2008zz}. For illustration purpose, we applied $n_2=2$ and $3$ for $\gamma A$ case.

As one can see, when the energy of photon is around $1.5 \, \textrm{GeV}$, the cross section can reach $12 \!\sim\! 24 \,\textrm{nb}$, which is sizable and might be measurable in the proposed experiment in this work. Consequently, experimental observation of $\phi$-meson produced by photon-nucleon interactions is imperative, providing an unambiguous signal for the gluonic probe of the SRC and impacting a range of measurements that rely on comprehension of nucleus structure.

The recent study has presented the first measurement of $J/\psi$ photoproduction from nuclei below the energy threshold of $8.2$\,GeV \cite{Pybus:2024ifi}. The data point corresponds to photon energies in the range $7\,\textrm{GeV} \leq E_\gamma \leq 8.2\,\textrm{GeV}$ (centered at $E_\gamma \sim 7.8\,\textrm{GeV}$), with a cross section of approximately $0.09\,\textrm{nb} \leq \sigma_{\gamma A}/Z \leq 0.18\,\textrm{nb}$. Notably, the energy fraction parameter method can yield reasonable agreement with this measurement ($0.04\,\textrm{nb} \leq \sigma_{\gamma A}/Z \leq 0.16\,\textrm{nb}$ \cite{Xu:2019wso}). As shown above, our model predicts a sub-threshold $\phi$-meson photoproduction cross section, which is over two orders of magnitude larger than the experimentally measured $J/\psi$ sub-threshold production. This also demonstrates the feasibility of observing this process in fixed-target experiments (such as LEPS, CLAS, as well as their upgrades \cite{Muramatsu:2021bpl,Burkert:2020akg}) We encourage experimentalists to pursue this investigation.

%%%%%%%%%%%%%%%%%%%%%%%%%%%%%%%%%%%%%%%%%%%%%%%%%%%%
\textit{The opportunities in BESIII and STCF.}---It would be beneficial to investigate this sub- and near-threshold $\phi$ production in electron-positron collision experiments. We will show that photons produced in BESIII and STCF can serve as a suitable platform for investigating nucleon structure. We will discuss the three- and two-body decays of $J/\psi (\psi(3686))$, along with the radiative Bhabha and $e^+ e^- \to \pi^+ \pi^- \gamma_{\text{ISR}}$ processes, as photon sources in order.

The BESIII detector records symmetric $e^+ e^-$ collisions at the BEPCII collider, providing a significant number of photons for study. These photons can interact with materials in the beam pipe alongside the $e^+ e^-$ beam, offering a unique opportunity to investigate SRCs. The materials in the beam pipe can be considered as targets of beryllium ($^9$Be), as detailed in \cite{BESIII:2020nme,Yuan:2021yks}. With a large production cross section in $e^+ e^-$ collisions, the BESIII detector has collected a sample of $1 \times 10^{10} \, J/\psi$ and $3 \times 10^{9} \, \psi(3686)$ events. Intense photons can be produced by the three-body decays $J/\psi (\psi(3686)) \to \gamma \pi \pi $ or $J/\psi (\psi(3686)) \to \gamma K K $, with energies ranging from $1.30$ to $1.85 \,\textrm{GeV}$, covering the entire range from sub-threshold to near-threshold photoproduction of the $\phi$ particle. The two-body decays such as $J/\psi \to \gamma \eta(\eta')$ and $\psi(3686) \to \gamma \eta(\eta')$ can provide us with signals of fixed-point photon energies. By tagging the particles recoiling against a photon in $J/\psi (\psi(3686))$ decays, one can precisely determine the photon's momentum and direction. Its interaction with the target material allows for precise measurements in particle and nuclear physics, including SRC pairs in nuclei. Additionally, the radiative Bhabha events $e^+ e^- \to \gamma e^+ e^-$, which serve as background in many BESIII studies, can also yield a large number of photons with suitable energy ($1.20 \sim 2.45 \,\textrm{GeV}$). Furthermore, a substantial number of photons with suitable energies can be obtained through $e^+ e^- \to \pi^+ \pi^- \gamma_{\text{ISR}}$. These photons, when produced, can interact with the beam pipe and have a probability of producing $\phi$ particles. By measuring the decay mode $\phi \to K^+ K^-$ and determining the interaction location, we can expect to obtain the shape of the photoproduction cross section of $\phi$ as a function of photon energy.

In the following analysis, we will examine each of these distinct photon sources sequentially.

\begin{itemize}
  \item {\it Photon source from three-body decays:}

      \quad In the three-body decays of $J/\psi(\psi(3686)) \to \gamma \pi \pi$, there are only two charged tracks originating from the interaction point which can be efficiently selected and identified. With known four-momenta of the initial $J/\psi(\psi(3686))$ and those of the $\pi^{\pm}$, the photon can be selected. The maximum momentum of the photon is $1.55$\,GeV for $J/\psi$ and $1.85$\,GeV for $\psi(3686)$, corresponding to the case when the two $\pi$'s fly in the same direction, opposite to the photon; the minimum momentum of the photon is zero, when the two $\pi$'s fly back-to-back with the same magnitude of the momentum. Therefore, the energy of photons covers the range of interest in SRC physics. The momentum spectrum of the photon is determined based on early published results on $J/\psi \to \gamma \pi \pi$ from BESIII \cite{BESIII:2015rug}. We employ $\rho_{J/\psi \to \gamma \pi \pi}(E_\gamma)$ to represent the energy distribution of photon numbers produced in the three-body decays $J/\psi \to \gamma \pi \pi$. A detailed description for this distribution is provided in the appendix. For the BESIII detector, the charged track resolution at 1\,GeV is 0.5\%. In our work, the photon energy is reconstructed from the recoil momentum of the $\pi^+ \pi^-$, yielding a photon energy resolution better than 10\,MeV. Since the photon energy range of interest (1.30$\sim$1.85\,GeV) is much larger than this resolution, the measurement will remain highly reliable.

      \quad A schematic diagram of $e^+ e^- \to J/\psi \to \gamma \pi^+ \pi^-$, followed by $\gamma$ interacting with a nucleon in the beam pipe to produce a $\phi$-meson, which subsequently decays to $K^+ K^-$, is illustrated in Fig.\,\ref{schematic}. The final state of the signal process is $\pi^+ \pi^- K^+ K^- p$, therefore the candidate events have five charged tracks with a positive net charge. It should be noted that, when the photon interacts with an $np$-SRC pair, both the paired proton and neutron will be ejected from the beam pipe. However, since BESIII cannot detect neutrons, the observable process remains as illustrated in Fig.\,\ref{schematic}.
      %%%%%%%%%%%%%%%%%%%%%%%
      \begin{figure}[htbp]
        %\begin{center}
        \includegraphics[width=1\columnwidth]{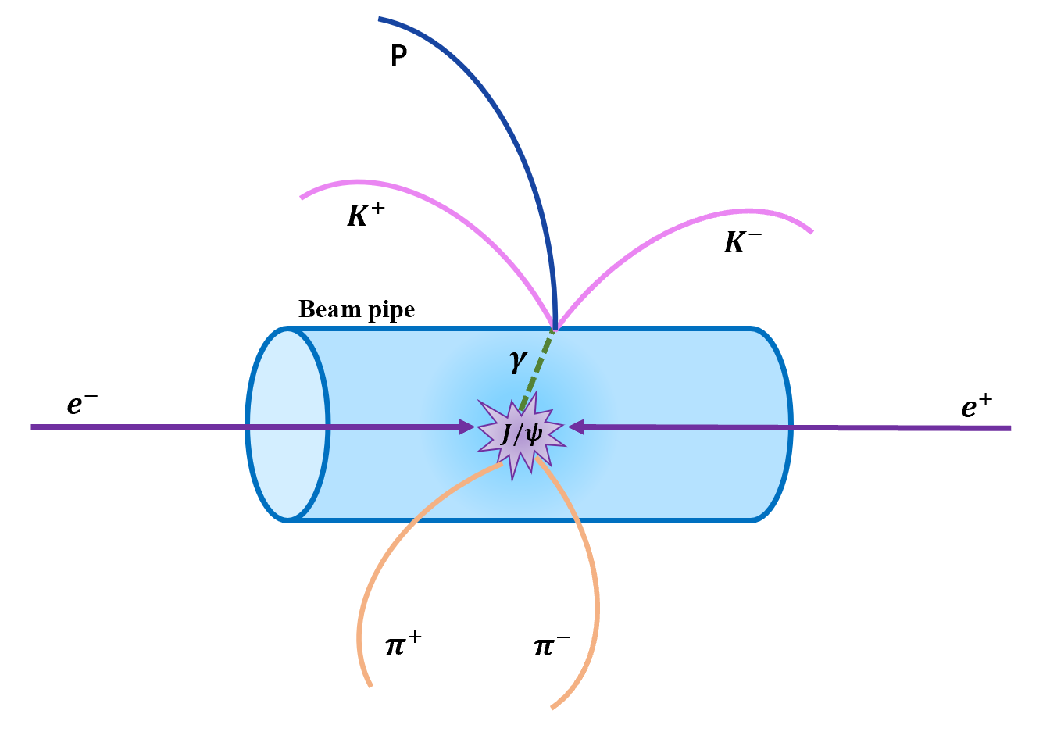}
        %\end{center}
        \caption{Schematic diagram of $e^+ e^- \to J/\psi \to \gamma \pi^+ \pi^-$, followed by $\gamma$ interacting with a proton in the material of beam pipe, $\gamma p \to \phi p \to K^+ K^- p$.}
        \label{schematic}
      \end{figure}
      %%%%%%%%%%%%%%%%%%%%%%%

  \item {\it Photon source from two-body decays:}

     \quad In the experiment, the two-body $J/\psi(\psi(3686)) \to \rho^{\pm} \pi^{\mp}, \rho^{0} \pi^{0}$ with the subsequent decay $\rho^{\pm} \to \pi^{\pm} \pi^0$ and $\pi^0 \to \gamma\gamma$ has been comprehensively investigated \cite{BES:2004mxa}, its invariant mass and angular distribution can be well-documented and the accuracy of the branching fractions can reach 0.8\%. This process could also be simulated using the available generators at BESIII, the $\gamma$ energy and $\pi-\pi$ invariant mass distributions exhibit consistency between the experimental data and Monte Carlo simulations, indicating its contributions can be estimated effectively. Therefore, the $J/\psi(\psi(3686)) \to \rho^{\pm} \pi^{\mp}, \rho^{0} \pi^{0}$ with the subsequent decay $\rho^{\pm} \to \pi^{\pm} \pi^0$ and $\pi^0 \to \gamma\gamma$ process can serve as a signal source for supplying the required photons. For similar reasons, other two-body decays such as $J/\psi \to \gamma \eta (\eta')$, $\psi(3686) \to \gamma \eta (\eta')$ can also be utilized as signal processes. They can help to enhance statistics in a photon energy range spanning around the threshold. Here, we take $\rho_{2\textrm{body}}(E_\gamma)$ to denote the energy distribution of photon numbers produced in these two-body decays. Relevant detailed evaluations are provided in the appendix.

  \item {\it Photon source from radiative Bhabha:}

     \quad For the radiative Bhabha process, we aim to select a single high-energy photon with energy exceeding 1.20\,GeV, accompanied by the electron and positron in the final state which will be recorded in the detector. To validate this method, we simulate the radiative Bhabha process using the Babayaga generator at $\sqrt{s}=3.773\,\textrm{GeV}$ \cite{CarloniCalame:2003yt}. The photon energy distribution is illustrated in Fig.\,\ref{energy-Bhabha}(a). In events where the photon energy exceeds 1.20\,GeV, we select the accompanying electron and positron, their energy distributions are presented in Fig.\,\ref{energy-Bhabha}(b). One can observe that the majority of electrons (positrons) accumulate around half of the center-of-mass energy. In order to mitigate the influence of multiple radiative photons, we can select electrons (positrons) with momenta exceeding 2.40\,GeV to identify the desired quasi-three-body process $\gamma e^+ e^-$. Consequently, the radiative Bhabha events can also serve as signal processes, yielding a large number of photons with suitable energy. We denote $\rho_{\textrm{Bha.}}(E_\gamma)$ to represent the energy distribution of photon numbers, its specific results can be found in the appendix.
     %%%%%%%%%%%%%%%%%%%%%%%
     \begin{figure}[htbp]
       \centering
       \includegraphics[width=1\columnwidth]{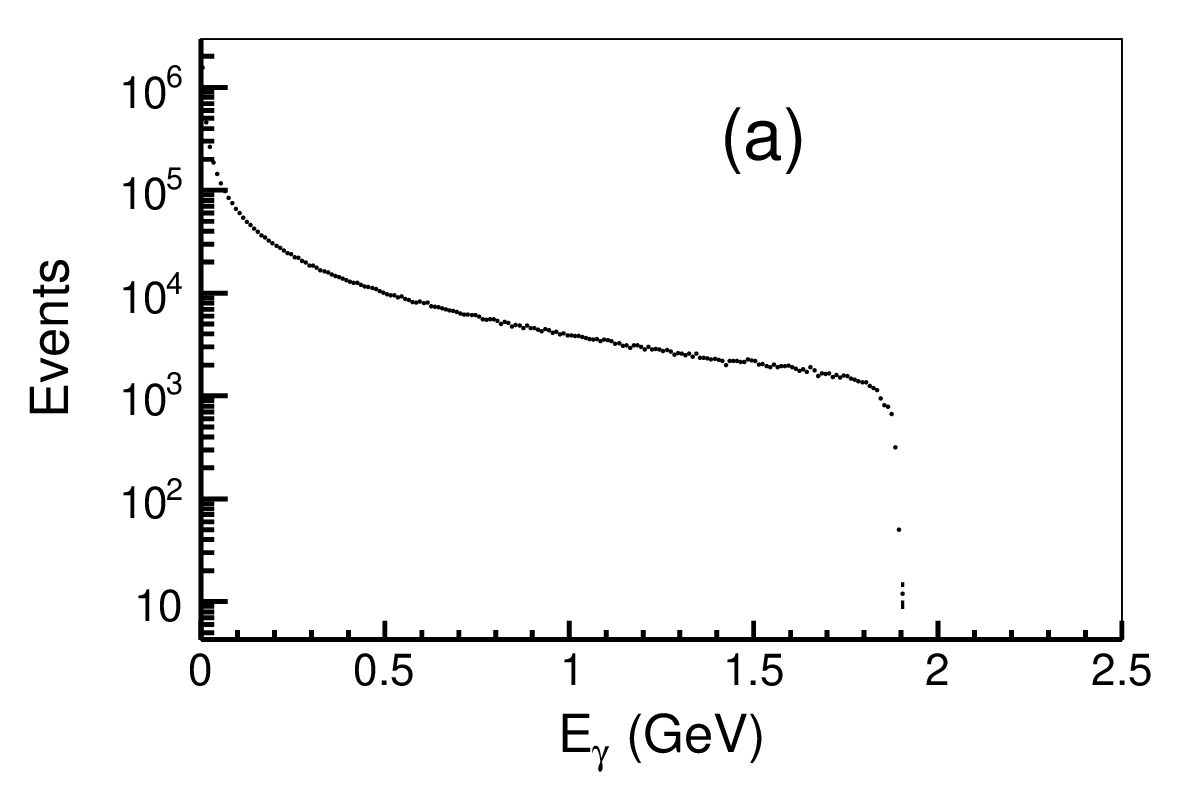}
       \centering
       \includegraphics[width=1\columnwidth]{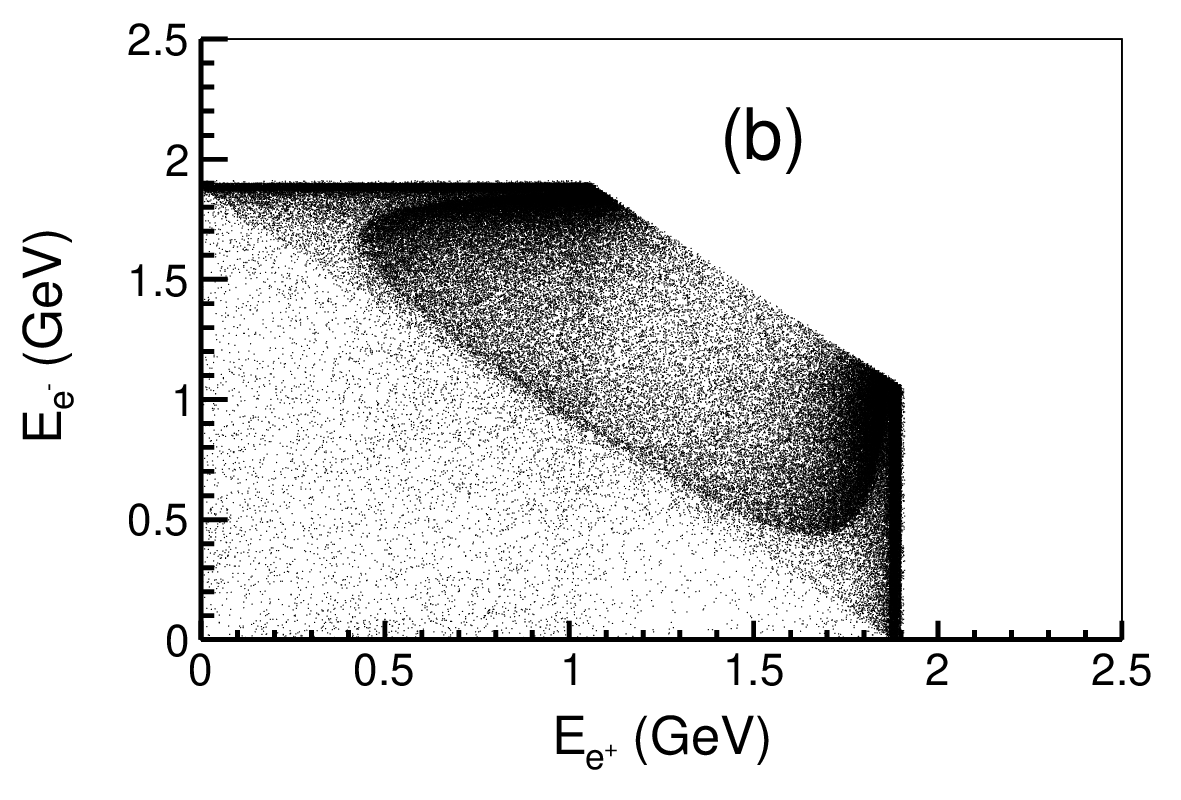}
       %\caption{fig2}
       \centering
       \caption{(a): The photon energy distribution of radiative Bhabha process at $\sqrt{s}=3.773\,\textrm{GeV}$. (b): The two dimensional energy distribution of $e^+$ versus $e^-$ with $E_\gamma>1.20 \, \textrm{GeV}$. These plots are obtained from the Monte Carlo truth information.}
       \label{energy-Bhabha}
     \end{figure}
     %%%%%%%%%%%%%%%%%%%%%%%

  \item {\it Photon source from $e^+ e^- \to \pi^+ \pi^- \gamma_{\textrm{ISR}}$:}

\quad Despite the generally smaller cross sections for $e^+e^- \to$ hadrons compared to Bhabha scattering, the specific process $e^+e^- \to \pi^+\pi^-\gamma_{\text{ISR}}$ is dominated by the resonant channel $e^+e^- \to \rho\gamma_{\text{ISR}}$. This results in a peak in the ISR photon energy distribution between 1.7 and 1.8 GeV (Fig.\,\ref{energy-pipi}), making it a useful photon source.
     \begin{figure}[htbp]
       \centering
       \includegraphics[width=1\columnwidth]{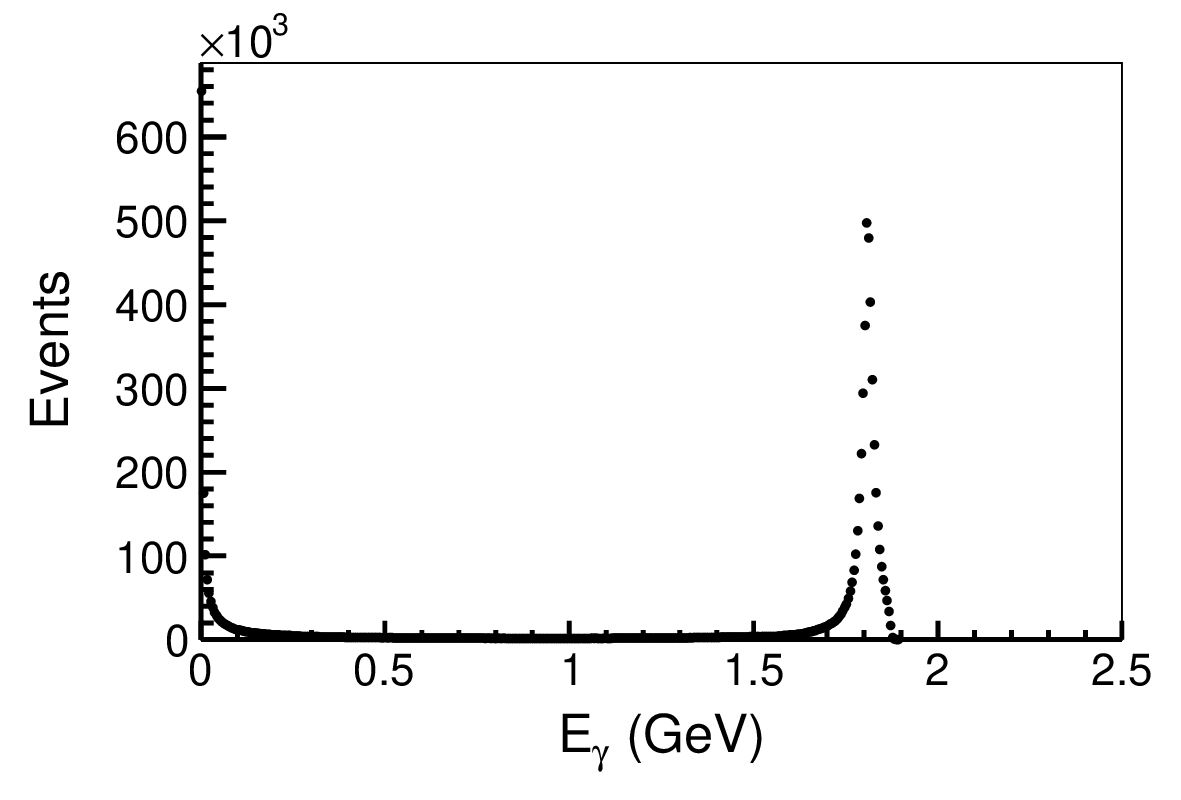}
       \centering
       \caption{The photon energy distribution of $e^+e^- \to \pi^+\pi^-\gamma_{\text{ISR}}$ at $\sqrt{s}=3.773\,\textrm{GeV}$. This plot is obtained from the Monte Carlo truth information.}
       \label{energy-pipi}
     \end{figure}

     \quad The energy of the leading order ISR photons is shown in Fig.\,\ref{energy-pipi}. This distribution has its peak  between 1.7 and 1.8 GeV. We use $\rho_{\pi\pi\gamma_{\textrm{ISR}}}(E_\gamma)$ to denote its energy distribution of photon numbers, which is collected in the appendix.

\end{itemize}

Fig.\,\ref{PhotonMomentum}(a) shows the energy spectrums of the photons from different processes discussed above and Fig.\,\ref{PhotonMomentum}(b) presents the sum of all these channels. As one can seen, a large number of photons ranging from the sub- to near-threshold can be generated, serving as the bullets for subsequent target shooting.

%%%%%%%%%%%%%%%%%%%%%%%
\begin{figure}[htbp]
\centering
\includegraphics[width=1\columnwidth]{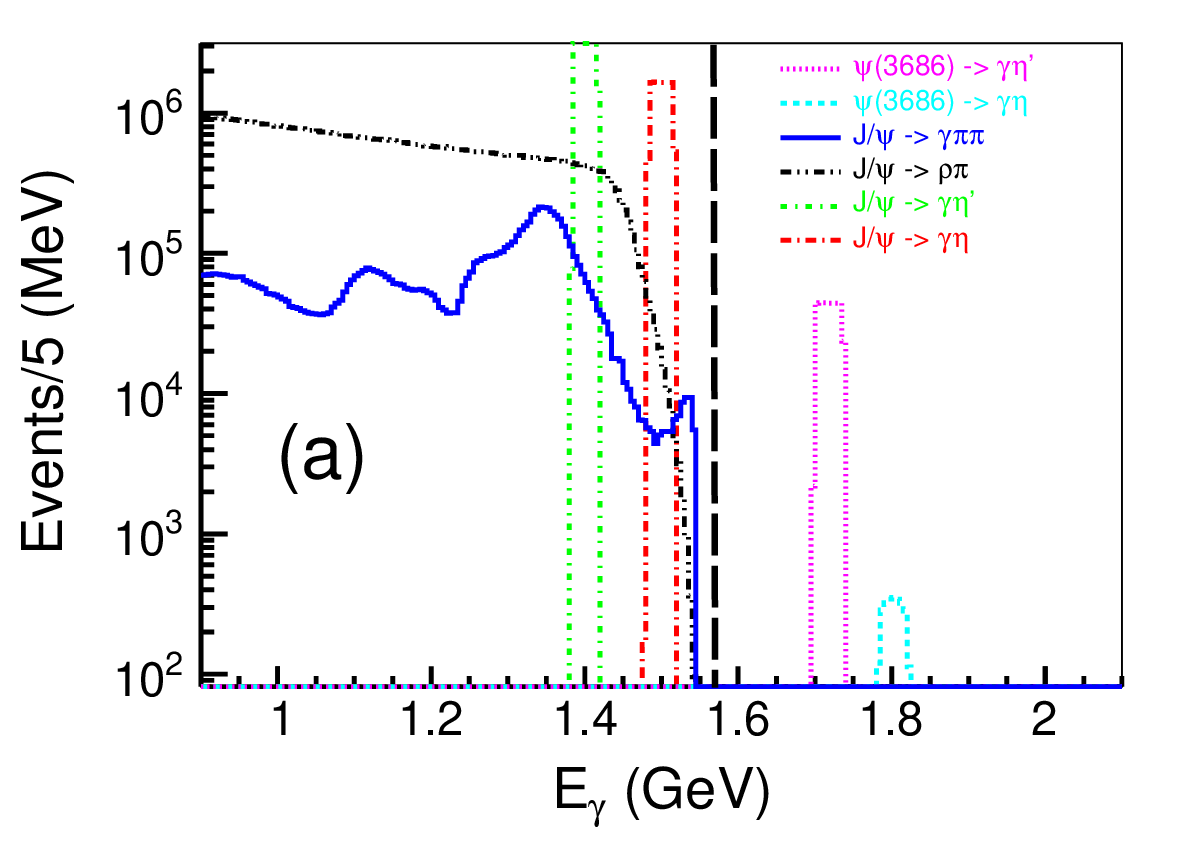}
\includegraphics[width=1\columnwidth]{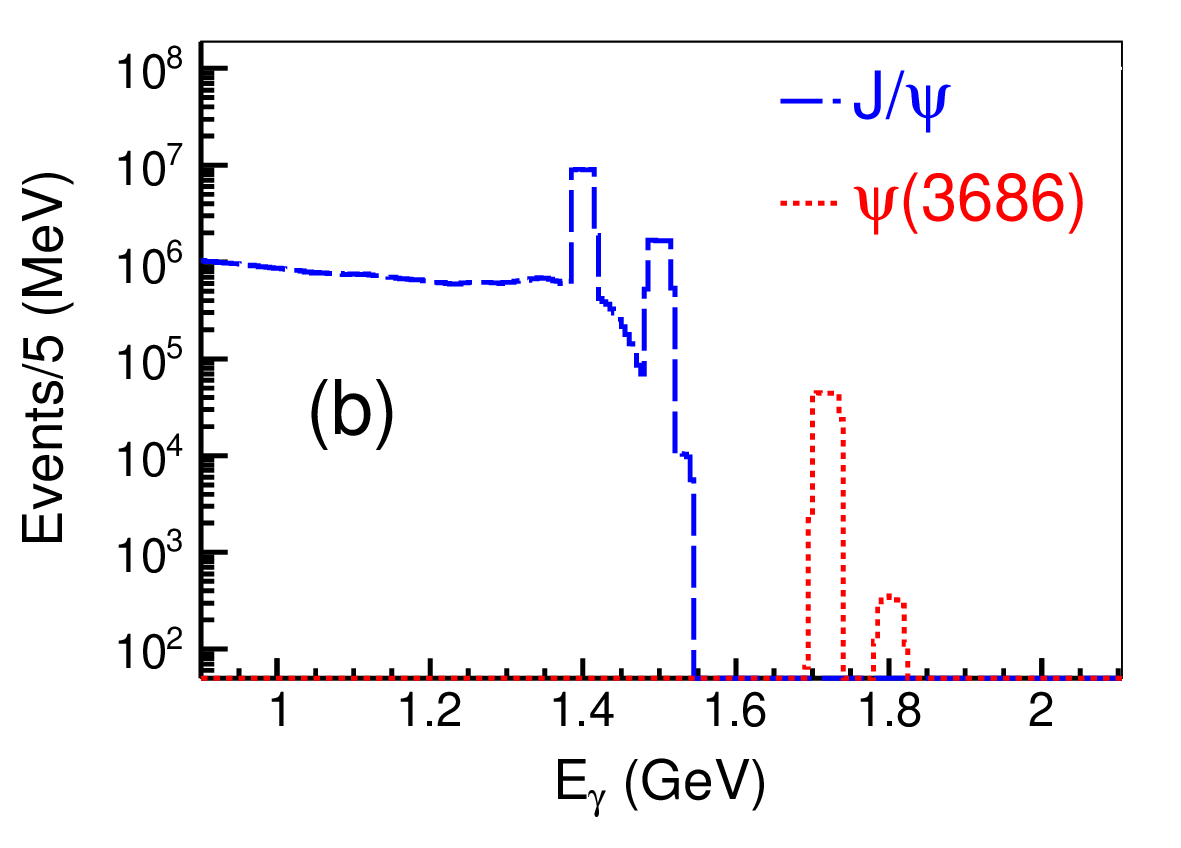}
%\caption{fig1}
\centering
\caption{(a): The photon enegy spectrums from the three-body decays $J/\psi \to \gamma \pi \pi$, as well as two-body decays $J/\psi \to \rho \pi$, $J/\psi (\psi(3686)) \to \gamma \eta (\eta')$. These processes offer photons ranging from below the threshold energy to above (the bold dashed line represents the threshold energy $E_\gamma \!=\! 1.57$\,GeV). (b): The photon energy spectrums from combined $J/\psi$ and $\psi(3686)$ decay channels. These plots are obtained from the Monte Carlo truth information.}
\label{PhotonMomentum}
\end{figure}
%%%%%%%%%%%%%%%%%%%%%%%

In the actual experimental data analysis, various background channels must be carefully considered in the analysis, and several selection criteria can effectively suppress these backgrounds:
\begin{enumerate}
  \item {\it Vertex constraint of $K^+ K^-$:}

     \quad The final state $K^+ K^-$ pair in our signal process originates from the beam pipe rather than the interaction point (IP).  By reconstructing the production vertex of the $K^+ K^-$, we can effectively suppress backgrounds such as $J/\psi \to \phi \pi^+ \pi^- \to K^+ K^- \pi^+ \pi^-$, where the kaons are produced directly at the IP.

  \item {\it Baryon number conservation:}

     \quad Because our signal process appears to violate baryon number conservation, with the proton in the final state originating from the beam pipe rather than the interaction point, we can suppress baryon-number-conserving backgrounds by reconstructing all charged tracks.

  \item {\it Photon four-momentum reconstruction:}

     \quad Taking the three-body decay $J/\psi \to \gamma \pi \pi$ as an example, the photon four-momentum is calculated using missing momentum:
     \begin{eqnarray}
       p_\gamma = p_{J/\psi} - p_{\pi^+} - p_{\pi^-} \,.
     \end{eqnarray}
     The invariant mass of the photon should peak near zero, which helps reject backgrounds with additional charged tracks (e.g., $J/\psi \to p \bar p \pi^+ \pi^-$).
\end{enumerate}

These criteria provide a robust and efficient way to suppress background channels. However, some background leakage is inevitable, and its rate may ultimately limit the statistical precision of the measurement. We will leave the more comprehensive background analysis to future experimental work.

The signal events of the reaction between photons and the nucleons in beam pipe can be estimated using \cite{BESIII:2023clq,Dai:2022wpg,BESIII:2024geh}
\begin{eqnarray}\label{events_estimate_1}
  && N^{\textrm{sig}} = N^{\textrm{sig}}_{\textrm{sub}} + N^{\textrm{sig}}_{\textrm{near}} \nn\\
  && \quad = \mathcal{B}(\phi \to K^+ K^-) \Big( \int_{1.30\,\textrm{GeV}}^{1.57\,\textrm{GeV}} dE_\gamma \, \sigma_{\gamma A \to \phi}(E_{\gamma}) \, \mathcal{L}_{\textrm{eff}}(E_\gamma) \nn\\
  &&\quad + \int_{1.57\,\textrm{GeV}}^{1.85\,\textrm{GeV}} dE_\gamma \, \sigma_{\gamma A \to \phi}(E_{\gamma}) \, \mathcal{L}_{\textrm{eff}}(E_\gamma) \Big) \,.
\end{eqnarray}
Here, signal events are classified as $N^{\textrm{sig}}_{\textrm{sub}}$ or $N^{\textrm{sig}}_{\textrm{near}}$, depending on the incident photon energy. The branching fraction $\mathcal{B}(\phi \to K^+ K^-)=(49.1\pm 0.5)\%$. The cross section $\sigma_{\gamma A \to \phi}(E_{\gamma})$ has been calculated in Eqs.\,(8) and (\ref{energyregion2}).

$\mathcal{L}_{\textrm{eff}}$ is the effective differential luminosity of the photon flux and target materials,
\begin{eqnarray}\label{effectiveL}
  \mathcal{L}_{\textrm{eff}}(E_\gamma) &=& \Big( \rho_{J/\psi \to \gamma \pi \pi}(E_\gamma) + \rho_{2\textrm{body}}(E_\gamma) +  \rho_{\textrm{Bha.}}(E_\gamma) \nn\\
  &&+\rho_{\pi\pi\gamma_{\textrm{ISR}}}(E_\gamma) \Big) \times \int_a^b N(x) \, dx \,.
\end{eqnarray}
The beam pipe consists of multiple layers of composite material, $N(x)$ is the number of nuclei per unit volume, $a=3.15\,\textrm{cm}$ and $b=3.37\,\textrm{cm}$ are the distances from the inner surface and outer surface of the beam pipe to the $e^+ e^-$ collision axis (the angular distribution of photons is not considered here). By integrating the material density along the beam pipe thickness, we have
\begin{eqnarray}\label{effN}
  \int_a^b N(x) \, dx = 2.1 \times 10^{22} \,\rm{cm}^{-2} \,.
\end{eqnarray}
More relevant details can be found in \cite{BESIII:2009fln}. Taking Eqs.\,(\ref{energyregion1}), (\ref{energyregion2}), (\ref{effectiveL}) and (\ref{effN}) into Eq.\,(\ref{events_estimate_1}), we obtain
\begin{eqnarray}\label{eventnumbers1}
  N^{\textrm{sig}} = 6.96 \,,
\end{eqnarray}
with
\begin{eqnarray}\label{eventnumbers2}
  N^{\textrm{sig}}_{\textrm{sub}} = 0.82 \,, \qquad N^{\textrm{sig}}_{\textrm{near}} = 6.14 \,.
\end{eqnarray}
It can be seen that the estimated number of signal events is rather limited. It would be quite challenging for BESIII to detect these signals experimentally, particularly for sub-threshold events. Nevertheless, these results indicate the potential of BESIII in elucidating this fundamental physics of nucleon structure. Taking into account additional possible processes as photon sources would lead to an increase in the estimated event count.

The STCF will reach a peak luminosity of $1\times 10^{35} \, \rm{cm}^{-2}\rm{s}^{-1}$, which is 100 times of the present BEPCII. Simply rescaling the numbers presented in Eq.\,(\ref{eventnumbers2}), we expect to obtain approximately one hundred sub-threshold events and several hundred near-threshold events. This makes the STCF a suitable machine for investigating the physics discussed in this work.

\textit{Summary.}---Significant progress has been achieved in the study of nucleon-nucleon interactions over the past three decades, particularly in the understanding of SRCs that occur when nucleons are closely overlapped. Numerous compelling results have emerged from experiments conducted at SLAC, BNL, and JLab. In this context, we have identified a promising avenue to advance the frontiers of QCD. We find that the conventional fixed-target experiments are well-suited for detecting the sub-threshold $\phi$-meson photoproduction. Moreover, our proposal involves leveraging future electron-positron experiments such as STCF to investigate nucleon structure through sub- and near-threshold $\phi$-meson photoproduction. The decays of $J/\psi$ and $\psi(3686)$, along with radiative Bhabha and $e^+ e^- \to \pi^+ \pi^- \gamma_{\text{ISR}}$ events, can serve as abundant sources of photons for this innovative study in nuclear and particle physics.

%%%%%%%%%%%%%%%%%%%
%\section{Acknowledgements}
%\label{Acknowledgements}
%%%%%%%%%%%%%%%%%%%
\textit{Acknowledgements.}---We thank Profs. Guang-Shun Huang, Shuang-Shi Fang, and Wen-Biao Yan for the inspiring discussions and suggestions on the measurements at BESIII. This work is supported in part by National Natural Science Foundation of China under Grant No. 12125503, 12335003, 12305106, 12475098, 12205255, and 12105247.

\begin{widetext}
\appendix
\section{The energy distribution of photon numbers}
The photon numbers for each process is calculated by considering corresponding initial state radiation, luminosities, cross sections, the branching fractons and total $J/\psi$ and $\psi(3686)$ events number. After taking these factors into consideration, we conducted a Monte Carlo truth investigation. Tab.\,\ref{rhodistribution} collects the energy distribution of photon numbers for the processes discussed above. It can be observed that the radiative Bhabha serves as the primary photon source, whereas the remaining processes make complementary contributions within certain energy ranges. All these analyses were performed assuming an integrated luminosity of $20\,\textrm{fb}^{-1}$ at $\sqrt{s}=3.773\,\textrm{GeV}$. However, BESIII has now collected the data samples wih an integrated luminosity of approximately $37.65 \,\textrm{fb}^{-1}$ above 3.5 GeV. This larger dataset yields a factor of two increase in the number of signal events.

For the two-body decays \(J/\psi \rightarrow \gamma \eta\), \(J/\psi \rightarrow \gamma \eta'\), \(\psi(3686) \rightarrow \gamma \eta\), and \(\psi(3686) \rightarrow \gamma \eta'\), the photon energy spectrum is given by \(E_{\gamma} = (M^2 - m^2) / (2M)\), where \(M\) and \(m\) are the masses of the initial charmonium state and the final-state meson, respectively. The event number for each process is calculated by the product of the corresponding branching fractions~\cite{ParticleDataGroup:2024cfk} and the total number of \(J/\psi\) or \(\psi(3686)\)~\cite{BESIII:2016kpv,BESIII:2024lks}.

For the three-body decay \(J/\psi \rightarrow \gamma \pi \pi\), the photon momentum spectrum is taken from early BESIII results~\cite{BESIII:2015rug}. The simulated sample for the process is normalized using the corresponding branching fractions~\cite{ParticleDataGroup:2024cfk} and the total number of \(J/\psi\) events~\cite{BESIII:2016kpv}.

The $\pi^+\pi^-\gamma_{\text{ISR}}$ and radiative Bhabha are generated using \textsc{PHOKHARA}~\cite{Rodrigo:2001kf} and \textsc{Babayaga3.5}~\cite{CarloniCalame:2000pz}, respectively, with normalization by the corresponding cross sections and the integrated luminosity at \(\sqrt{s}=3.773\,\textrm{GeV}\)~\cite{BESIII:2024lbn}.

The $\rho_{2\textrm{body}}(E_\gamma)$ in Eq.\,(\ref{effectiveL}) is the sum of several two-body decay processes:
\begin{eqnarray}
  \rho_{2\textrm{body}}(E_\gamma) = \rho_{J/\psi \to \rho \pi}(E_\gamma) + \rho_{J/\psi \to \gamma \eta}(E_\gamma) + \rho_{J/\psi \to \gamma \eta'}(E_\gamma) + \rho_{\psi(3686) \to \gamma \eta}(E_\gamma) + \rho_{\psi(3686) \to \gamma \eta'}(E_\gamma) \,.
\end{eqnarray}

\begin{table}[!htbp]
	\centering
	\renewcommand{\arraystretch}{1.5}
	\caption{ The energy distribution of photon numbers $\rho(E_\gamma)$ for various processes. }\label{rhodistribution}
	\begin{tabular}{c||c |c |c |c |c |c |c |c}
		\hline\hline
		 ~~~~~$E_\gamma$~~~~~ & ~~~~~$\rho_{\textrm{Bha.}}$~~~~~ & ~~~~$\rho_{\pi\pi\gamma_{\textrm{ISR}}}$~~~~ & ~~~$\rho_{J/\psi \to \gamma \pi \pi}$~~~ & ~~~$\rho_{J/\psi \to \rho \pi}$~~~ & ~~$\rho_{J/\psi \to \gamma \eta}$~~ & ~~$\rho_{J/\psi \to \gamma \eta'}$~~ & ~~$\rho_{\psi(3686) \to \gamma \eta}$~~ & ~~$\rho_{\psi(3686) \to \gamma \eta'}$~~ \\
		\hline
		$1.30\,\textrm{GeV}$ & $5.50 \times 10^{7}$ & $8.01 \times 10^{4}$ & $1.11 \times 10^{6}$ & $5.02 \times 10^{6}$ & $0$                  & $0$
& $0$                  & $0$                \\
		$1.35\,\textrm{GeV}$ & $5.25 \times 10^{7}$ & $8.96 \times 10^{4}$ & $1.91 \times 10^{6}$ & $4.67 \times 10^{6}$ & $0$                  & $0$                       & $0$                  & $0$    \\
		$1.40\,\textrm{GeV}$ & $4.80 \times 10^{7}$ & $1.03 \times 10^{5}$ & $8.50 \times 10^{5}$ & $4.19 \times 10^{6}$ & $0$              & $5.28 \times 10^{7}$ & $0$                  & $0$                    \\
		$1.45\,\textrm{GeV}$ & $4.64 \times 10^{7}$ & $1.16 \times 10^{5}$ & $1.87 \times 10^{5}$ & $2.22 \times 10^{6}$ & $0$                  & $0$                  & $0$                  & $0$   \\
		$1.50\,\textrm{GeV}$ & $4.68 \times 10^{7}$ & $1.28 \times 10^{5}$ & $5.60 \times 10^{4}$ & $2.37 \times 10^{5}$ & $1.09 \times 10^{7}$ & $0$                 & $0$                  & $0$   \\
		$1.55\,\textrm{GeV}$ & $4.22 \times 10^{7}$ & $1.44 \times 10^{5}$ & $3.97 \times 10^{4}$ & $3.17 \times 10^{3}$ & $0$                  & $0$                         & $0$                  & $0$     \\
		$1.60\,\textrm{GeV}$ & $4.08 \times 10^{7}$ & $1.91 \times 10^{5}$ & $0$                  & $0$                  & $0$                  & $0$                       & $0$                  & $0$   \\
        $1.65\,\textrm{GeV}$ & $3.83 \times 10^{7}$ & $3.17 \times 10^{5}$ & $0$                  & $0$                  & $0$                  & $0$
& $0$                  & $0$   \\
		$1.70\,\textrm{GeV}$ & $3.49 \times 10^{7}$ & $5.96 \times 10^{5}$ & $0$                  & $0$                  & $0$                  & $0$                  & $0$                  & $1.12 \times 10^{5}$      \\
		$1.75\,\textrm{GeV}$ & $3.27 \times 10^{7}$ & $1.59 \times 10^{6}$ & $0$                  & $0$                  & $0$                  & $0$                 & $0$                  & $5.59 \times 10^{4}$    \\
		$1.80\,\textrm{GeV}$ & $2.88 \times 10^{7}$ & $1.04 \times 10^{7}$ & $0$                  & $0$                  & $0$                  & $0$                    & $2.48 \times 10^{3}$ & $1.67 \times 10^{5}$     \\
		$1.85\,\textrm{GeV}$ & $1.99 \times 10^{7}$ & $3.36 \times 10^{6}$ & $0$                  & $0$                  & $0$                  & $0$                      & $0$                  & $0$     \\
		\hline\hline
	\end{tabular}
\end{table}

\end{widetext}


\begin{thebibliography}{}

%\cite{Frankfurt:1988nt}
\bibitem{Frankfurt:1988nt}
L.~L.~Frankfurt and M.~I.~Strikman,
%``Hard Nuclear Processes and Microscopic Nuclear Structure,''
Phys. Rept. \textbf{160}, 235-427 (1988)
doi:10.1016/0370-1573(88)90179-2
%744 citations counted in INSPIRE as of 23 Jun 2024

%\cite{Hen:2016kwk}
\bibitem{Hen:2016kwk}
O.~Hen, G.~A.~Miller, E.~Piasetzky and L.~B.~Weinstein,
%``Nucleon-Nucleon Correlations, Short-lived Excitations, and the Quarks Within,''
Rev. Mod. Phys. \textbf{89}, no.4, 045002 (2017)
doi:10.1103/RevModPhys.89.045002
[arXiv:1611.09748 [nucl-ex]].
%294 citations counted in INSPIRE as of 26 Dec 2023

%\cite{Arrington:2022sov}
\bibitem{Arrington:2022sov}
J.~Arrington, N.~Fomin and A.~Schmidt,
%``Progress in understanding short-range structure in nuclei: an experimental perspective,''
Ann. Rev. Nucl. Part. Sci. \textbf{72}, 307-337 (2022)
doi:10.1146/annurev-nucl-102020-022253
[arXiv:2203.02608 [nucl-ex]].
%32 citations counted in INSPIRE as of 23 Jun 2024

%\cite{Frankfurt:1993sp}
\bibitem{Frankfurt:1993sp}
L.~L.~Frankfurt, M.~I.~Strikman, D.~B.~Day and M.~Sargsian,
%``Evidence for short range correlations from high Q**2 (e, e-prime) reactions,''
Phys. Rev. C \textbf{48}, 2451-2461 (1993)
doi:10.1103/PhysRevC.48.2451
%217 citations counted in INSPIRE as of 23 Jun 2024

%\cite{CLAS:2005ola}
\bibitem{CLAS:2005ola}
K.~S.~Egiyan \textit{et al.} [CLAS],
%``Measurement of 2- and 3-nucleon short range correlation probabilities in nuclei,''
Phys. Rev. Lett. \textbf{96}, 082501 (2006)
doi:10.1103/PhysRevLett.96.082501
[arXiv:nucl-ex/0508026 [nucl-ex]].
%312 citations counted in INSPIRE as of 24 Jun 2024

%\cite{Fomin:2011ng}
\bibitem{Fomin:2011ng}
N.~Fomin, J.~Arrington, R.~Asaturyan, F.~Benmokhtar, W.~Boeglin, P.~Bosted, A.~Bruell, M.~H.~S.~Bukhari, M.~E.~Christy and E.~Chudakov, \textit{et al.}
%``New measurements of high-momentum nucleons and short-range structures in nuclei,''
Phys. Rev. Lett. \textbf{108}, 092502 (2012)
doi:10.1103/PhysRevLett.108.092502
[arXiv:1107.3583 [nucl-ex]].
%196 citations counted in INSPIRE as of 24 Jun 2024

%\cite{EuropeanMuon:1983wih}
\bibitem{EuropeanMuon:1983wih}
J.~J.~Aubert \textit{et al.} [European Muon],
%``The ratio of the nucleon structure functions $F2_n$ for iron and deuterium,''
Phys. Lett. B \textbf{123}, 275-278 (1983)
doi:10.1016/0370-2693(83)90437-9
%1569 citations counted in INSPIRE as of 26 Dec 2023

%\cite{Weinstein:2010rt}
\bibitem{Weinstein:2010rt}
L.~B.~Weinstein, E.~Piasetzky, D.~W.~Higinbotham, J.~Gomez, O.~Hen and R.~Shneor,
%``Short Range Correlations and the EMC Effect,''
Phys. Rev. Lett. \textbf{106}, 052301 (2011)
doi:10.1103/PhysRevLett.106.052301
[arXiv:1009.5666 [hep-ph]].
%219 citations counted in INSPIRE as of 26 Dec 2023

%\cite{Hen:2012fm}
\bibitem{Hen:2012fm}
O.~Hen, E.~Piasetzky and L.~B.~Weinstein,
%``New data strengthen the connection between Short Range Correlations and the EMC effect,''
Phys. Rev. C \textbf{85}, 047301 (2012)
doi:10.1103/PhysRevC.85.047301
[arXiv:1202.3452 [nucl-ex]].
%96 citations counted in INSPIRE as of 26 Dec 2023

%\cite{Aschenauer:2017oxs}
\bibitem{Aschenauer:2017oxs}
E.~C.~Aschenauer, S.~Fazio, M.~A.~C.~Lamont, H.~Paukkunen and P.~Zurita,
%``Nuclear Structure Functions at a Future Electron-Ion Collider,''
Phys. Rev. D \textbf{96}, no.11, 114005 (2017)
doi:10.1103/PhysRevD.96.114005
[arXiv:1708.05654 [nucl-ex]].
%68 citations counted in INSPIRE as of 24 Jun 2024

%\cite{Xu:2019wso}
\bibitem{Xu:2019wso}
J.~Xu and F.~Yuan,
%``Gluonic Probe for the Short Range Correlation in Nucleus,''
Phys. Lett. B \textbf{801}, 135187 (2020)
doi:10.1016/j.physletb.2019.135187
[arXiv:1908.10413 [hep-ph]].
%8 citations counted in INSPIRE as of 02 Jan 2024

%\cite{Hatta:2019ocp}
\bibitem{Hatta:2019ocp}
Y.~Hatta, M.~Strikman, J.~Xu and F.~Yuan,
%``Sub-threshold $J/\psi$ and $\Upsilon$ production in $\gamma A$ collisions,''
Phys. Lett. B \textbf{803}, 135321 (2020)
doi:10.1016/j.physletb.2020.135321
[arXiv:1911.11706 [hep-ph]].
%16 citations counted in INSPIRE as of 02 Jan 2024

%\cite{Sun:2021pyw}
\bibitem{Sun:2021pyw}
P.~Sun, X.~B.~Tong and F.~Yuan,
%``Near threshold heavy quarkonium photoproduction at large momentum transfer,''
Phys. Rev. D \textbf{105}, no.5, 054032 (2022)
doi:10.1103/PhysRevD.105.054032
[arXiv:2111.07034 [hep-ph]].
%35 citations counted in INSPIRE as of 02 Jul 2025

%\cite{Wang:2024cpx}
\bibitem{Wang:2024cpx}
W.~Wang, J.~Xu, X.~H.~Yang and S.~Zhao,
%``Linear relation between short range correlation and EMC effect of gluons in nuclei,''
Eur. Phys. J. A \textbf{61}, no.5, 112 (2025)
doi:10.1140/epja/s10050-025-01588-4
[arXiv:2401.16662 [hep-ph]].
%1 citations counted in INSPIRE as of 06 Aug 2025

%\cite{Bosted:2008mn}
\bibitem{Bosted:2008mn}
P.~Bosted, J.~Dunne, C.~A.~Lee, P.~Junnarkar, M.~Strikman, J.~Arrington, R.~Asaturyan, F.~Benmokhtar, M.~E.~Christy and E.~Chudakov, \textit{et al.}
%``Search for Sub-threshold Photoproduction of J/Psi Mesons,''
Phys. Rev. C \textbf{79}, 015209 (2009)
doi:10.1103/PhysRevC.79.015209
[arXiv:0809.2284 [nucl-ex]].
%12 citations counted in INSPIRE as of 24 Jun 2024

%\cite{Pybus:2024ifi}
\bibitem{Pybus:2024ifi}
J.~R.~Pybus, L.~Ehinger, T.~Kolar, B.~Devkota, P.~Sharp, B.~Yu, M.~M.~Dalton, D.~Dutta, H.~Gao and O.~Hen, \textit{et al.}
%``First Measurement of Near-Threshold and Subthreshold J/{\ensuremath{\psi}} Photoproduction off Nuclei,''
Phys. Rev. Lett. \textbf{134}, no.20, 201903 (2025)
doi:10.1103/PhysRevLett.134.201903
[arXiv:2409.18463 [nucl-ex]].
%6 citations counted in INSPIRE as of 28 Aug 2025

%\cite{GlueX:2019mkq}
\bibitem{GlueX:2019mkq}
A.~Ali \textit{et al.} [GlueX],
%``First Measurement of Near-Threshold J/\ensuremath{\psi} Exclusive Photoproduction off the Proton,''
Phys. Rev. Lett. \textbf{123}, no.7, 072001 (2019)
doi:10.1103/PhysRevLett.123.072001
[arXiv:1905.10811 [nucl-ex]].
%194 citations counted in INSPIRE as of 24 Jun 2024

%\cite{GlueX:2023pev}
\bibitem{GlueX:2023pev}
S.~Adhikari \textit{et al.} [GlueX],
%``Measurement of the J/$\psi $ photoproduction cross section over the full near-threshold kinematic region,''
Phys. Rev. C \textbf{108}, no.2, 025201 (2023)
doi:10.1103/PhysRevC.108.025201
[arXiv:2304.03845 [nucl-ex]].
%37 citations counted in INSPIRE as of 24 Jun 2024

%\cite{Hatta:2025vhs}
\bibitem{Hatta:2025vhs}
Y.~Hatta, H.~T.~Klest, K.~Passek-K. and J.~Schoenleber,
%``Deeply virtual $?$-meson production near threshold,''
[arXiv:2501.12343 [hep-ph]].
%8 citations counted in INSPIRE as of 30 Oct 2025

%\cite{Qian:2010rr}
\bibitem{Qian:2010rr}
X.~Qian, W.~Chen, H.~Gao, K.~Hicks, K.~Kramer, J.~M.~Laget, T.~Mibe, Y.~Qiang, S.~Stepanyan and D.~J.~Tedeschi, \textit{et al.}
%``Near-threshold Photoproduction of Phi Mesons from Deuterium,''
Phys. Lett. B \textbf{696}, 338-342 (2011)
doi:10.1016/j.physletb.2010.12.065
[arXiv:1011.1305 [nucl-ex]].
%15 citations counted in INSPIRE as of 25 Jun 2024

%\cite{Wang:2022uch}
\bibitem{Wang:2022uch}
X.~Y.~Wang, C.~Dong and Q.~Wang,
%``Mass radius and mechanical properties of the proton via strange \ensuremath{\phi} meson photoproduction,''
Phys. Rev. D \textbf{106}, no.5, 056027 (2022)
doi:10.1103/PhysRevD.106.056027
[arXiv:2206.11644 [nucl-th]].
%8 citations counted in INSPIRE as of 25 Jun 2024

%\cite{Gao:2000az}
\bibitem{Gao:2000az}
H.~Gao, T.~S.~H.~Lee and V.~Marinov,
%``Phi0 - N bound state,''
Phys. Rev. C \textbf{63}, 022201 (2001)
doi:10.1103/PhysRevC.63.022201
[arXiv:nucl-th/0010042 [nucl-th]].
%66 citations counted in INSPIRE as of 02 Jul 2025

%\cite{Sekihara:2010rw}
\bibitem{Sekihara:2010rw}
T.~Sekihara, A.~Martinez Torres, D.~Jido and E.~Oset,
%``Theoretical study of incoherent $\phi$ photoproduction on a deuteron target,''
Eur. Phys. J. A \textbf{48}, 10 (2012)
doi:10.1140/epja/i2012-12010-1
[arXiv:1008.4422 [nucl-th]].
%5 citations counted in INSPIRE as of 02 Jul 2025

%\cite{Aachen-Berlin-Bonn-Hamburg-Heidelberg-Munich:1968rzt}
\bibitem{Aachen-Berlin-Bonn-Hamburg-Heidelberg-Munich:1968rzt}
 [Aachen-Berlin-Bonn-Hamburg-Heidelberg-Munich],
%``Photoproduction of Meson and Baryon Resonances at Energies up to 5.8 GEV,''
Phys. Rev. \textbf{175}, 1669-1696 (1968)
doi:10.1103/PhysRev.175.1669
%404 citations counted in INSPIRE as of 25 Jun 2024

%\cite{Mcclellan:1971tk}
\bibitem{Mcclellan:1971tk}
G.~Mcclellan, N.~B.~Mistry, P.~Mostek, H.~Ogren, A.~Osborne, J.~Swartz, R.~Talman and G.~Diambrini-Palazzi,
%``Photoproduction of phi0 mesons from hydrogen, deuterium, and complex nuclei,''
Phys. Rev. Lett. \textbf{26}, 1593-1596 (1971)
doi:10.1103/PhysRevLett.26.1593
%49 citations counted in INSPIRE as of 25 Jun 2024

%\cite{Anderson:1972ac}
\bibitem{Anderson:1972ac}
R.~L.~Anderson, B.~Gottschalk, D.~Gustavson, D.~Ritson, G.~A.~Weitsch, B.~H.~Wiik, H.~J.~Halpern, R.~Prepost and D.~H.~Tompkins,
%``MEASUREMENTS OF THE REACTION gamma P ---\ensuremath{>} phi P WITH UNPOLARIZED BREMSSTRAHLUNG,''
Phys. Rev. Lett. \textbf{30}, 149 (1973)
doi:10.1103/PhysRevLett.30.149
%33 citations counted in INSPIRE as of 25 Jun 2024

%\cite{Egloff:1979mg}
\bibitem{Egloff:1979mg}
R.~M.~Egloff, P.~J.~Davis, G.~Luste, J.~F.~Martin, J.~D.~Prentice, D.~O.~Caldwell, J.~P.~Cumalat, A.~M.~Eisner, A.~Lu and R.~J.~Morrison, \textit{et al.}
%``Measurements of Elastic Rho and Phi Meson Photoproduction Cross-Sections on Protons from 30 GeV to 180 GeV,''
Phys. Rev. Lett. \textbf{43}, 657 (1979)
doi:10.1103/PhysRevLett.43.657
%104 citations counted in INSPIRE as of 25 Jun 2024

%\cite{ZEUS:1999ptu}
\bibitem{ZEUS:1999ptu}
J.~Breitweg \textit{et al.} [ZEUS],
%``Measurement of diffractive photoproduction of vector mesons at large momentum transfer at HERA,''
Eur. Phys. J. C \textbf{14}, 213-238 (2000)
doi:10.1007/s100520000374
[arXiv:hep-ex/9910038 [hep-ex]].
%133 citations counted in INSPIRE as of 25 Jun 2024

%\cite{LEPS:2005hax}
\bibitem{LEPS:2005hax}
T.~Mibe \textit{et al.} [LEPS],
%``Diffractive phi-meson photoproduction on proton near threshold,''
Phys. Rev. Lett. \textbf{95}, 182001 (2005)
doi:10.1103/PhysRevLett.95.182001
[arXiv:nucl-ex/0506015 [nucl-ex]].
%133 citations counted in INSPIRE as of 25 Jun 2024

%\cite{Dey:2014tfa}
\bibitem{Dey:2014tfa}
B.~Dey \textit{et al.} [CLAS],
%``Data analysis techniques, differential cross sections, and spin density matrix elements for the reaction $\gamma p \rightarrow \phi p$,''
Phys. Rev. C \textbf{89}, no.5, 055208 (2014)
doi:10.1103/PhysRevC.89.055208
[arXiv:1403.2110 [nucl-ex]].
%79 citations counted in INSPIRE as of 25 Jun 2024

%\cite{LEPS:2009nuw}
\bibitem{LEPS:2009nuw}
W.~C.~Chang \textit{et al.} [LEPS],
%``Measurement of the incoherent $\gamma d \to \phi p n$ photoproduction near threshold,''
Phys. Lett. B \textbf{684}, 6-10 (2010)
doi:10.1016/j.physletb.2009.12.051
[arXiv:0907.1705 [nucl-ex]].
%28 citations counted in INSPIRE as of 25 Jun 2024

%\cite{CLAS:2007xhu}
\bibitem{CLAS:2007xhu}
T.~Mibe \textit{et al.} [CLAS],
%``First measurement of coherent phi-meson photoproduction on deuteron at low energies,''
Phys. Rev. C \textbf{76}, 052202 (2007)
doi:10.1103/PhysRevC.76.052202
[arXiv:nucl-ex/0703013 [nucl-ex]].
%40 citations counted in INSPIRE as of 30 Oct 2025

%\cite{Chang:2007fc}
\bibitem{Chang:2007fc}
W.~C.~Chang, K.~Horie, S.~Shimizu, M.~Miyabe, D.~S.~Ahn, J.~K.~Ahn, H.~Akimune, Y.~Asano, S.~Date and H.~Ejiri, \textit{et al.}
%``Forward coherent phi-meson photoproduction from deuterons near threshold,''
Phys. Lett. B \textbf{658}, 209-215 (2008)
doi:10.1016/j.physletb.2007.11.009
[arXiv:nucl-ex/0703034 [nucl-ex]].
%43 citations counted in INSPIRE as of 30 Oct 2025

%\cite{BESIII:2009fln}
\bibitem{BESIII:2009fln}
M.~Ablikim \textit{et al.} [BESIII],
%``Design and Construction of the BESIII Detector,''
Nucl. Instrum. Meth. A \textbf{614}, 345-399 (2010)
doi:10.1016/j.nima.2009.12.050
[arXiv:0911.4960 [physics.ins-det]].
%1183 citations counted in INSPIRE as of 25 Jun 2024

%\cite{Achasov:2023gey}
\bibitem{Achasov:2023gey}
M.~Achasov, X.~C.~Ai, R.~Aliberti, L.~P.~An, Q.~An, X.~Z.~Bai, Y.~Bai, O.~Bakina, A.~Barnyakov and V.~Blinov, \textit{et al.}
%``STCF conceptual design report (Volume 1): Physics \& detector,''
Front. Phys. (Beijing) \textbf{19}, no.1, 14701 (2024)
doi:10.1007/s11467-023-1333-z
[arXiv:2303.15790 [hep-ex]].
%58 citations counted in INSPIRE as of 25 Jun 2024

%\cite{Brodsky:2000zc}
\bibitem{Brodsky:2000zc}
S.~J.~Brodsky, E.~Chudakov, P.~Hoyer and J.~M.~Laget,
%``Photoproduction of charm near threshold,''
Phys. Lett. B \textbf{498}, 23-28 (2001)
doi:10.1016/S0370-2693(00)01373-3
[arXiv:hep-ph/0010343 [hep-ph]].
%116 citations counted in INSPIRE as of 26 Jun 2024

%\cite{Subedi:2008zz}
\bibitem{Subedi:2008zz}
R.~Subedi, R.~Shneor, P.~Monaghan, B.~D.~Anderson, K.~Aniol, J.~Annand, J.~Arrington, H.~Benaoum, W.~Bertozzi and F.~Benmokhtar, \textit{et al.}
%``Probing Cold Dense Nuclear Matter,''
Science \textbf{320}, 1476-1478 (2008)
doi:10.1126/science.1156675
[arXiv:0908.1514 [nucl-ex]].
%438 citations counted in INSPIRE as of 07 Jun 2024

%\cite{Muramatsu:2021bpl}
\bibitem{Muramatsu:2021bpl}
N.~Muramatsu, M.~Yosoi, T.~Yorita, Y.~Ohashi, J.~K.~Ahn, S.~Ajimura, Y.~Asano, W.~C.~Chang, J.~Y.~Chen and S.~Dat'e, \textit{et al.}
%``SPring-8 LEPS2 beamline: A facility to produce a multi-GeV photon beam via laser Compton scattering,''
Nucl. Instrum. Meth. A \textbf{1033}, 166677 (2022)
doi:10.1016/j.nima.2022.166677
[arXiv:2112.07832 [physics.acc-ph]].
%18 citations counted in INSPIRE as of 30 Oct 2025

%\cite{Burkert:2020akg}
\bibitem{Burkert:2020akg}
V.~D.~Burkert, L.~Elouadrhiri, K.~P.~Adhikari, S.~Adhikari, M.~J.~Amaryan, D.~Anderson, G.~Angelini, M.~Antonioli, H.~Atac and S.~Aune, \textit{et al.}
%``The CLAS12 Spectrometer at Jefferson Laboratory,''
Nucl. Instrum. Meth. A \textbf{959}, 163419 (2020)
doi:10.1016/j.nima.2020.163419
%148 citations counted in INSPIRE as of 30 Oct 2025

%\cite{BESIII:2020nme}
\bibitem{BESIII:2020nme}
M.~Ablikim \textit{et al.} [BESIII],
%``Future Physics Programme of BESIII,''
Chin. Phys. C \textbf{44}, no.4, 040001 (2020)
doi:10.1088/1674-1137/44/4/040001
[arXiv:1912.05983 [hep-ex]].
%501 citations counted in INSPIRE as of 25 Jun 2024

%\cite{Yuan:2021yks}
\bibitem{Yuan:2021yks}
C.~Z.~Yuan and M.~Karliner,
%``Cornucopia of Antineutrons and Hyperons from a Super J/\ensuremath{\psi} Factory for Next-Generation Nuclear and Particle Physics High-Precision Experiments,''
Phys. Rev. Lett. \textbf{127}, no.1, 012003 (2021)
doi:10.1103/PhysRevLett.127.012003
[arXiv:2103.06658 [hep-ex]].
%19 citations counted in INSPIRE as of 29 Jun 2024

%\cite{BESIII:2015rug}
\bibitem{BESIII:2015rug}
M.~Ablikim \textit{et al.} [BESIII],
%``Amplitude analysis of the $\pi^{0}\pi^{0}$~system produced in radiative $J/\psi$~decays,''
Phys. Rev. D \textbf{92}, no.5, 052003 (2015)
[erratum: Phys. Rev. D \textbf{93}, no.3, 039906 (2016)]
doi:10.1103/PhysRevD.92.052003
[arXiv:1506.00546 [hep-ex]].
%76 citations counted in INSPIRE as of 31 Aug 2024

%\cite{BES:2004mxa}
\bibitem{BES:2004mxa}
J.~Z.~Bai \textit{et al.} [BES],
%``Measurement of the branching fraction of J / psi ---\ensuremath{>} pi+ pi- pi0,''
Phys. Rev. D \textbf{70}, 012005 (2004)
doi:10.1103/PhysRevD.70.012005
[arXiv:hep-ex/0402013 [hep-ex]].
%76 citations counted in INSPIRE as of 16 Jan 2025

%\cite{CarloniCalame:2003yt}
\bibitem{CarloniCalame:2003yt}
C.~M.~Carloni Calame, G.~Montagna, O.~Nicrosini and F.~Piccinini,
%``The BABAYAGA event generator,''
Nucl. Phys. B Proc. Suppl. \textbf{131}, 48-55 (2004)
doi:10.1016/j.nuclphysbps.2004.02.008
[arXiv:hep-ph/0312014 [hep-ph]].
%129 citations counted in INSPIRE as of 16 Jan 2025

%\cite{Dai:2022wpg}
\bibitem{Dai:2022wpg}
J.~Dai, H.~B.~Li, H.~Miao and J.~Zhang,
%``Prospects to study hyperon-nucleon interactions at BESIII*,''
Chin. Phys. C \textbf{48}, no.7, 073003 (2024)
doi:10.1088/1674-1137/ad3dde
[arXiv:2209.12601 [hep-ex]].
%9 citations counted in INSPIRE as of 30 Aug 2024

%\cite{BESIII:2023clq}
\bibitem{BESIII:2023clq}
M.~Ablikim \textit{et al.} [BESIII],
%``First Study of Reaction \ensuremath{\Xi}0n\textrightarrow{}\ensuremath{\Xi}-p Using \ensuremath{\Xi}0-Nucleus Scattering at an Electron-Positron Collider,''
Phys. Rev. Lett. \textbf{130}, no.25, 251902 (2023)
doi:10.1103/PhysRevLett.130.251902
[arXiv:2304.13921 [hep-ex]].
%9 citations counted in INSPIRE as of 01 Jul 2024

%\cite{BESIII:2024geh}
\bibitem{BESIII:2024geh}
M.~Ablikim \textit{et al.} [BESIII],
%``First Study of Antihyperon-Nucleon Scattering \ensuremath{\Lambda}\textasciimacron{}p\textrightarrow{}\ensuremath{\Lambda}\textasciimacron{}p and Measurement of \ensuremath{\Lambda}p\textrightarrow{}\ensuremath{\Lambda}p Cross Section,''
Phys. Rev. Lett. \textbf{132}, no.23, 231902 (2024)
doi:10.1103/PhysRevLett.132.231902
[arXiv:2401.09012 [hep-ex]].
%4 citations counted in INSPIRE as of 30 Aug 2024

%\cite{ParticleDataGroup:2024cfk}
\bibitem{ParticleDataGroup:2024cfk}
S.~Navas \textit{et al.} [Particle Data Group],
%``Review of particle physics,''
Phys. Rev. D \textbf{110}, no.3, 030001 (2024)
doi:10.1103/PhysRevD.110.030001
%3081 citations counted in INSPIRE as of 05 Nov 2025

%\cite{BESIII:2016kpv}
\bibitem{BESIII:2016kpv}
M.~Ablikim \textit{et al.} [BESIII],
%``Determination of the number of $J/\psi$ events with inclusive $J/\psi$ decays,''
Chin. Phys. C \textbf{41}, no.1, 013001 (2017)
doi:10.1088/1674-1137/41/1/013001
[arXiv:1607.00738 [hep-ex]].
%91 citations counted in INSPIRE as of 05 Nov 2025

%\cite{BESIII:2024lks}
\bibitem{BESIII:2024lks}
M.~Ablikim \textit{et al.} [BESIII],
%``Determination of the number of {\ensuremath{\psi}}(3686) events taken at BESIII*,''
Chin. Phys. C \textbf{48}, no.9, 093001 (2024)
doi:10.1088/1674-1137/ad595b
[arXiv:2403.06766 [hep-ex]].
%65 citations counted in INSPIRE as of 05 Nov 2025

%\cite{Rodrigo:2001kf}
\bibitem{Rodrigo:2001kf}
G.~Rodrigo, H.~Czyz, J.~H.~Kuhn and M.~Szopa,
%``Radiative return at NLO and the measurement of the hadronic cross-section in electron positron annihilation,''
Eur. Phys. J. C \textbf{24}, 71-82 (2002);
%doi:10.1007/s100520200912
%[arXiv:hep-ph/0112184 [hep-ph]].
%239 citations counted in INSPIRE as of 02 Nov 2025
%\cite{Czyz:2008kw}
%\bibitem{Czyz:2008kw}
H.~Czyz, J.~H.~Kuhn and A.~Wapienik,
%``Four-pion production in tau decays and e+e- annihilation: An Update,''
Phys. Rev. D \textbf{77}, 114005 (2008);
%doi:10.1103/PhysRevD.77.114005
%[arXiv:0804.0359 [hep-ph]].
%58 citations counted in INSPIRE as of 02 Nov 2025
%\cite{Czyz:2009vj}
%\bibitem{Czyz:2009vj}
H.~Czyz and J.~H.~Kuhn,
%``Strong and Electromagnetic J/psi and psi(2S) Decays into Pion and Kaon Pairs,''
Phys. Rev. D \textbf{80}, 034035 (2009).
%doi:10.1103/PhysRevD.80.034035
%[arXiv:0904.0515 [hep-ph]].
%16 citations counted in INSPIRE as of 02 Nov 2025



%\cite{CarloniCalame:2000pz}
\bibitem{CarloniCalame:2000pz}
C.~M.~Carloni Calame, C.~Lunardini, G.~Montagna, O.~Nicrosini and F.~Piccinini,
%``Large angle Bhabha scattering and luminosity at flavor factories,''
Nucl. Phys. B \textbf{584}, 459-479 (2000);
%doi:10.1016/S0550-3213(00)00356-4
%[arXiv:hep-ph/0003268 [hep-ph]].
%199 citations counted in INSPIRE as of 02 Nov 2025
%\cite{CarloniCalame:2019nra}
%\bibitem{CarloniCalame:2019nra}
C.~M.~Carloni Calame, G.~Montagna, O.~Nicrosini and F.~Piccinini,
%``Status of the BabaYaga event generator,''
EPJ Web Conf. \textbf{218}, 07004 (2019).
%doi:10.1051/epjconf/201921807004
%10 citations counted in INSPIRE as of 02 Nov 2025



%\cite{BESIII:2024lbn}
\bibitem{BESIII:2024lbn}
M.~Ablikim \textit{et al.} [BESIII],
%``Measurement of integrated luminosity of data collected at 3.773 GeV by BESIII from 2021 to 2024*,''
Chin. Phys. C \textbf{48}, no.12, 123001 (2024).
%doi:10.1088/1674-1137/ad70a0
%[arXiv:2406.05827 [hep-ex]].
%47 citations counted in INSPIRE as of 02 Nov 2025



\end{thebibliography}
\end{document}